\documentstyle[aps,epsf]{revtex}                  
\begin{document}
\pagestyle{empty}                                      
\preprint{
\font\fortssbx=cmssbx10 scaled \magstep2
\hbox to \hsize{
\hfill$\raise .5cm\vtop{\hbox{NCKU-HEP-99-04}}$}}
\draft
\vfill
\title{PQCD analysis of exclusive $B$ meson decays
\footnote{talk presented at DPF99, UCLA, 1999}}
\vfill
\author{Hsiang-nan Li}

\address{Department of Physics, National Cheng-Kung University,
Tainan 701, Taiwan, Republic of China}

\vfill

\maketitle
\begin{abstract}

The perturbative QCD formalism for exclusive heavy meson decays, especially
the three-scale factorization theorem for nonleptonic modes is reviewed
and compared to the conventional Bauer-Stech-Wirbel model.

\end{abstract}

\vskip 1.0cm


\pagestyle{plain}

\section{INTRODUCTION}

Recently, perturbative QCD (PQCD) has been proposed to be an alternative
approach to the study of heavy hadron decays \cite{LY1,L1,LY3}, which
complements the heavy quark effective theory (HQET) \cite{G} for inclusive
processes and the Bauer-Stech-Wirbel (BSW) method \cite{BSW} for exclusive
processes. The basic idea is the factorization
theorem, which states that nonperturbative dynamics involved in a physical
quantity can be factorized into a hadron distribution function or
wave function. The remaining part, if characterized by a large scale, is
calculable in perturbation theory. The distribution (wave) function,
though not calculable, is universal. A physical quantity is then expressed
as the convolution of perturbative parts with nonperturbative distribution
(wave) functions. Once a distribution (wave) function
is determined, say, from experimental data of some processes, it can
be employed to make predictions of other processes involving the same
hadron.

Within the PQCD framework, we have been able to explained the semileptonic
branching ratio $B_{\rm SL}$ and the average charm yield $n_c$ in inclusive
$B$ meson decays simultaneously \cite{CCL}. Extending the same formalism to
inclusive $\Lambda_b$ baryon decays straightforwardly, we have predicted 
a low lifetime ratio $\tau(\Lambda_b)/\tau(B_d)$ \cite{CLY}, that the HQET
approach can not achieve. For exclusive $B$ meson decays, various
transition form factors at large recoil, and factorizable and
nonfactorrizable contributions can be evaluated systematically \cite{CL,LLG}.  
In this talk I will review the PQCD analysis of exclusive $B$ meson decays,
concentrating on the three-scale factorization theorem for nonleptonic
modes \cite{CL}, and explore the relation beetween the PQCD formalism
and the BSW model.

\section{THREE-SCALE FACTORIZATION THEOREMS}

Nonleptonic $B$ meson decays involve three scales: the $W$ boson mass $M_W$ 
(the matching scale), at which the matching conditions of the
effective Hamiltonian to the full Hamiltonian are defined, the hard
gluon momentum $t$ of order the $B$ meson mass $M_B$, which reflects the 
specific dynamics of different
decay modes, and the factorization scale $1/b$ of order the QCD
scale $\Lambda_{\rm QCD}$, $b$ being the conjugate variable of parton
transverse momenta. Dynamics below $1/b$ is regarded as being
completely nonperturbative, and parametrized into a meson wave fucntion
$\phi(x)$, $x$ being the momentum fraction. Dynamics above the scale
$1/b$ is perturbative, and absorbed into a hard subamplitude $H(t)$, if it
is characterized by $t$, and into a ''harder" function $H_r(M_W)$, if it
is characterized by $M_W$. Semeleptonic decays depend only on the scales
$t$ and $1/b$, and thus their factorization involves only hard 
subamplitudes and meson wave functions.

Radiative corrections produce two types of large logarithms $\ln(M_W/t)$
and $\ln(tb)$. The former are summed to give the Wilson evolution $C(t)$
from $M_W$ down to $t$, that connects the harder function and the hard
subamplitude. While the latter are summed to give the renormalization-group
(RG) evolution $\Gamma(t,b)$ from $t$ to $1/b$, that connects the hard
subamplitude and the meson wave functions. There exist also double
logarithms $\ln^2(Pb)$ from the overlap of collinear and soft divergences,
$P$ being the dominant light-cone component of a meson momentum. The
resummation of these double logarithms leads to an exponential 
$\exp[-s(P,b)]$, which suppresses the long-distance contributions in the 
large $b$
region, and improves the applicability of PQCD around the energy scale of
few GeV \cite{LS}. The $b$ quark mass scale is located in the range of
applicability. 

For the harder function, characterized by the large scale $M_W$,
we adopt its lowest-order expression $H_r(M_W)=H_r^{(0)}=1$. Because of
Sudakov suppression, the hard subamplitude $H(t)$ can be evaluated
reliably in perturbation theory. Note that $H(t)$ contains both factorizable
and nonfactorizable contributions, each of which includes the types of
external $W$-emissions, internal $W$-emissions, and $W$-exchanges. Take
lowest-order diagrams as an example. If a hard gluon attaches the
valence quarks of the same meson, the diagram gives factorizable
contribution. If a hard gluon attaches the valence quarks of different
mesons, the diagram gives nonfactorizable contribution. 
The Wilson coefficients $C(t)$ and the Sudakov factor 
$S(P,t,b)=\exp[-s(P,b)]\Gamma(t,b)$ for heavy meson decays up to the 
accuracy of next-to-leading
orders have been derived in \cite{REVIEW} and in \cite{LY1,L1}, respectively. 
Therefore, a three-scale factorization formula
possesses the typical expression, 
\begin{eqnarray}
C(t)\otimes H(t)\otimes \phi(x)\otimes S(P,t,b)\;,
\label{for}
\end{eqnarray} 
where all the convolution factors except the meson wave function $\phi$
are calculable.

Nonperturbative wave functions, though not calculable, are universal. They 
absorb the long-distance
dynamics of a decay process, which are insensitive to the short-distance
dynamics involved in the specific decay of the $b$ quark into
light quarks with large energy release. The universality of nonperturbative
wave functions is the fundamental concept of PQCD factorization theorems.
Because of the universality, the strategy of the PQCD approach is
as follows: evaluate all perturbative factors for some decay modes, and
adjust the wave functions such that predictions from the corresponding
factorization formulas match experimental data. At this stage, the
nonperturbative wave functions are determined up to the twists and orders of
the coupling constant, at which the factorization formulas are constructed.
Then evaluate the perturbative factors for another decay mode. Input
the extracted wave functions into the factorization formulas of the same
twist and orders, and make predictions. With this strategy, PQCD
factorization theorems are model independent and possess a predictive power.

\section{THE $B\to D^{(*)}\pi$ DECAYS}

We take the nonleptonic decays $B\to D^{(*)}\pi$ as an example of
the application of the three-scale factorization theorem.
The decay rates of $B\to D^{(*)}\pi$ have the expression
\begin{equation}
\Gamma_i=\frac{1}{128\pi}G_F^2|V_{cb}|^2|V_{ud}|^2M_B^3\frac{(1-r^2)^3}{r}
|{\cal M}_i|^2\;,
\label{dr}
\end{equation}
where $i=1$, 2, 3, and 4 denote the modes $B^-\to D^0\pi^-$,
${\bar B}^0\to D^+\pi^-$, $B^-\to D^{*0}\pi^-$, and
${\bar B}^0\to D^{*+}\pi^-$, respectively. The decay
amplitudes ${\cal M}_i$ are written as
\begin{eqnarray}
{\cal M}_1&=&f_\pi[(1+r)\xi_+-(1-r)\xi_-]+
f_D\xi_{\rm int}+{\cal M}_{\rm ext}+{\cal M}_{\rm int}\;,
\label{M1}\\
{\cal M}_2&=&f_\pi[(1+r)\xi_+-(1-r)\xi_-]+f_B
\xi_{\rm exc}+{\cal M}_{\rm ext}+{\cal M}_{\rm exc}\;,
\label{M2}\\
{\cal M}_3&=&\frac{1+r}{2r}f_\pi[(1+r)\xi_{A_1}-(1-r)(r\xi_{A_2}
+\xi_{A_3})]
+f_{D^*}\xi^*_{\rm int}+{\cal M}^*_{\rm ext}+{\cal M}^{*}_{\rm int}\;,
\label{M3}\\
{\cal M}_4&=&\frac{1+r}{2r}f_\pi[(1+r)\xi_{A_1}-(1-r)(r\xi_{A_2}
+\xi_{A_3})]
+f_B\xi^*_{\rm exc}+{\cal M}^*_{\rm ext}+{\cal M}^{*}_{\rm exc}\;,
\label{M4}
\end{eqnarray}
where $f_B$, $f_{D^{(*)}}$, and $f_\pi$ are the $B$ meson, $D^{(*)}$ meson,
and pion decay constants, respectively. The form factors $\xi_i$, $i=+$,
$-$, $V$, $A_1$, $A_2$, and $A_3$, denote the factorizable external
$W$-emission contributions. The form factors $\xi^{(*)}_{\rm int}$ and
$\xi^{(*)}_{\rm exc}$ denote the factorizable internal $W$-emission and
$W$-exchange contributions, respectively.
The amplitudes ${\cal M}_{\rm ext}^{(*)}$,
${\cal M}_{\rm int}^{(*)}$, and ${\cal M}_{\rm exc}^{(*)}$ represent the
nonfactorizable external $W$-emission, internal $W$-emission, and
$W$-exchange contributions, respectively.

The momenta of the $B$ and $D^{(*)}$ mesons in light-cone coordinates are 
written as $P_1=(M_B/\sqrt{2})(1,1,{\bf 0}_T)$ and
$P_2=(M_B/\sqrt{2})(1,r^2,{\bf 0}_T)$, respectively, with $r=M_{D^{(*)}}/M_B$.
We define the momenta of light valence quarks in the $B$ and $D^{(*)}$
mesons as $k_1$ and $k_2$, respectively. $k_1$ has a minus component
$k_1^-$, giving the momentum fraction $x_1=k_1^-/P_1^-$, and small
transverse components ${\bf k}_{1T}$. $k_2$ has a large plus component
$k_2^+$, giving $x_2=k_2^+/P_2^+$, and small ${\bf k}_{2T}$. The pion
momentum is then $P_3=P_1-P_2$, whose nonvanishing component is only
$P_3^-$.

The resummation of the large logarithmic corrections to the 
meson wave functions leads to the exponentials,
\begin{eqnarray}
S_B(t)&=&\exp\left[-s(x_1P_1^-,b_1)
-2\int_{1/b_1}^{t}\frac{d{\bar \mu}}{\bar \mu}
\gamma(\alpha_s({\bar \mu}))\right]\;,
\label{sbb} \\
S_{D^(*)}(t)&=&\exp\left[-s(x_2P_2^+,b_2)-s((1-x_2)P_2^+,b_2)
-2\int_{1/b_2}^{t}
\frac{d{\bar \mu}}{\bar \mu}\gamma(\alpha_s({\bar \mu}))\right]\;,
\label{sbk}\\
S_\pi(t)&=&\exp\left[-s(x_3P_3^-,b_3)-s((1-x_3)P_3^-,b_3)-
2\int_{1/b_3}^{t}\frac{d{\bar \mu}}{\bar \mu}
\gamma(\alpha_s({\bar \mu}))\right]\;.
\label{wpe}
\end{eqnarray}
The variable $b_i$, $i=1$, 2, and 3, conjugate to the parton transverse 
momentum $k_{iT}$, represents the transverse extent of the corresponding
mesons. The exponential with the exponent involving the anomalous dimension
$\gamma=-\alpha_s/\pi$ corresponds to the RG evolution $\Gamma$ mentioned
above. For the explicit expression of the exponent $s$, refer to \cite{CS,BS}.

We present only the factorization formulas for the nonfactorizable 
amplitudes here. Those for factorizable contributions are referred to
\cite{LM}. The integration over $b_3$
can be performed trivially, leading to $b_3=b_1$ or $b_3=b_2$. Their
expressions are
\begin{eqnarray}
{\cal M}^{(*)}_{\rm ext}&=& 32\pi\sqrt{2N} C_F\sqrt{r}M_B^2
\int_0^1 [dx]\int_0^{\infty}
b_1 db_1 b_2 db_2
\phi_B(x_1)\phi_{D^{(*)}}(x_2)\phi_\pi(x_3)
\frac{C_2(t_b)}{N}S(t_b)|_{b_2=b_1,b_3=b_2}
\nonumber \\
& &\times\alpha_s(t_b)\biggl\{
[x_3(1-r^2)-x_1-\zeta^{(*)}_bx_2(r-r^2)]h^{(1)}_b(x_i,b_i)
-[x_3(1-r^2)-x_1+x_2]h^{(2)}_b(x_i,b_i) \biggr\}\;,
\label{mb}\\
{\cal M}^{(*)}_{\rm int}&=& 32\pi\sqrt{2N} C_F\sqrt{r}M_B^2
\int_0^1 [dx]\int_0^{\infty}b_1 db_1 b_2 db_2
\phi_B(x_1)\phi_{D^{(*)}}(x_2)\phi_\pi(x_3)
\frac{C_1(t_d)}{N}S(t_d)|_{b_3=b_1}
\nonumber \\
& &\times 
\alpha_s(t_d)\biggl\{[x_1-x_2-x_3(1-r^2)]h^{(1)}_d(x_i,b_i)
-[(x_1+x_2)(1+\zeta^{(*)}_d r^2)-1]h^{(2)}_d(x_i,b_i)
\biggr\}\;,
\label{md}\\
{\cal M}^{(*)}_{\rm exc}&=& 32 \pi\sqrt{2N} C_F\sqrt{r}M_B^2
\int_0^1 [dx]\int_0^{\infty}b_1 db_1 b_2 db_2
\phi_B(x_1)\phi_{D^{(*)}}(x_2)\phi_\pi(x_3)
\frac{C_1(t_f)}{N}S(t_f)|_{b_3=b_2}
\nonumber \\
& &\times
\alpha_s(t_f)\biggl\{
[x_3(1-r^2)-\zeta^{(*)}_f(x_1-x_2)r^2]h^{(1)}_f(x_i,b_i)
-[(x_1+x_2)(1+\zeta^{(*)}_fr^2)
-\zeta^{(*)}_fr^2]h^{(2)}_f(x_i,b_i) \biggr\}\;,
\label{mf}
\end{eqnarray}
with the number of color $N=3$, the color factor $C_F=4/3$,
the definition $[dx]\equiv dx_1dx_2dx_3$, and the constants
$\zeta_{b,d,f}=-\zeta^{*}_{b,d,f}=1$. The Wilson coefficients $C_{1,2}$
will be defined later. The complete Sudakov factor
is given by the product of Eqs.~(\ref{sbb})-(\ref{wpe}),
$S=S_BS_{D^{(*)}}S_\pi$.

The functions $h^{(j)}$, $j=1$ and 2, appearing in
Eqs.~(\ref{mb})-(\ref{mf}), are written as
\begin{eqnarray}
\everymath{\displaystyle}
h^{(j)}_b&=& \left[\theta(b_1-b_2)K_0\left(BM_B
b_1\right)I_0\left(BM_Bb_2\right)
+\theta(b_2-b_1)K_0\left(BM_B b_2\right)
I_0\left(BM_B b_1\right)\right]\;  \nonumber \\
&  & \times \left( \begin{array}{cc}
 K_{0}(B_{j}M_Bb_{2}) &  \mbox{for $B^2_{j} \geq 0$}  \\
 \frac{i\pi}{2} H_{0}^{(1)}(\sqrt{|B_{j}^2|}M_Bb_{2})  &
 \mbox{for $B^2_{j} \leq 0$}
  \end{array} \right)\;,          
\\
\everymath{\displaystyle}
h^{(j)}_d&=& \left[\theta(b_1-b_2)K_0\left(DM_B
b_1\right)I_0\left(DM_Bb_2\right)
+\theta(b_2-b_1)K_0\left(DM_B b_2\right)
I_0\left(DM_B b_1\right)\right]\;  \nonumber \\
&  & \times \left( \begin{array}{cc}
 K_{0}(D_{j}M_Bb_{2}) &  \mbox{for $D^2_{j} \geq 0$}  \\
 \frac{i\pi}{2} H_{0}^{(1)}(\sqrt{|D_{j}^2|}M_Bb_{2})  &
 \mbox{for $D^2_{j} \leq 0$}
  \end{array} \right)\;,          
\label{hjd}\\
\everymath{\displaystyle}
h^{(j)}_f&=& i\frac{\pi}{2}
\left[\theta(b_1-b_2)H_0^{(1)}\left(FM_B
b_1\right)J_0\left(FM_Bb_2\right)
+\theta(b_2-b_1)H_0^{(1)}\left(FM_B b_2\right)
J_0\left(FM_B b_1\right)\right]\;  \nonumber \\
&  & \times \left( \begin{array}{cc}
 K_{0}(F_{j}M_Bb_{1}) &  \mbox{for $F^2_{j} \geq 0$}  \\
 \frac{i\pi}{2} H_{0}^{(1)}(\sqrt{|F_{j}^2|}M_Bb_{1})  &
 \mbox{for $F^2_{j} \leq 0$}
  \end{array} \right)\;,          
\end{eqnarray}
with the variables
\begin{eqnarray}
B^{2}&=&x_{1}x_{2}\;,
\nonumber \\
B_{1}^{2}&=&x_{1}x_{2}-x_2x_{3}(1-r^{2})\;,
\nonumber \\
B_{2}^{2}&=&x_{1}x_{2}(1+r^{2})-(x_{1}-x_{2})(1-x_{3})(1-r^{2})\;,
\nonumber \\
D^{2}&=&x_{1}x_{3}(1-r^{2})\;,
\nonumber \\
D_{1}^{2}&=&F_1^2=(x_{1}-x_{2})x_{3}(1-r^{2})\;,
\nonumber \\
D_{2}^{2}&=&(x_{1}+x_{2})r^{2}-(1-x_{1}-x_{2})x_{3}(1-r^{2})\;,
\nonumber \\
F^{2}&=&x_{2}x_{3}(1-r^{2})\;,
\nonumber \\
F_{2}^{2}&=&x_{1}+x_{2}+(1-x_{1}-x_{2})x_{3}(1-r^{2})\;.
\end{eqnarray}
The scales $t^{(j)}$ are chosen as
\begin{eqnarray}
t_b&=&{\rm max}(BM_B,\sqrt{|B_1^2|}M_B,\sqrt{|B_2^2|}M_B,
1/b_1,1/b_2)\;,
\nonumber \\
t_d&=&{\rm max}(DM_B,\sqrt{|D_1^2|}M_B,\sqrt{|D_2^2|}M_B,
1/b_1,1/b_2)\;,
\nonumber \\
t_f&=&{\rm max}(FM_B,\sqrt{|F_1^2|}M_B,\sqrt{|F_2^2|}M_B,
1/b_1,1/b_2)\;.
\end{eqnarray}
The wave functions $\phi_i(x)$, $i=B$, $D^{(*)}$, and $\pi$,
satisfy the normalization
\begin{equation}
\int_0^1\phi_i(x)dx=\frac{f_i}{2\sqrt{6}}\;,
\label{dco}
\end{equation}
with the corresponding decay constants $f_i$.
It is easy to observe that Eqs.~(\ref{mb})-(\ref{mf}) agree with the
typical expression in Eq.~(\ref{for}).

\section{COMPARISION TO THE BSW METHOD}

In this section I compare the PQCD approach with the BSW model. Nonleptonic 
heavy meson decays occur through the Hamiltonian,
\begin{eqnarray}
H=\frac{G_F}{\sqrt 2}V_{ij}V_{kl}^{*}({\bar q}_l q_k)
({\bar q}_j q_i)\;,
\label{full}
\end{eqnarray}
where $G_F$ is the Fermi coupling constant, $V_{cb}$ the
Cabibbo-Kabayashi-Maskawa (CKM) matrix element, and
$({\bar q} q)={\bar q} \gamma_\mu(1-\gamma_5)q$ the $V-A$ current.
Hard gluon corrections cause an operator mixing, and their
RG summation leads to the effective Hamiltonian,
\begin{eqnarray}
H_{\rm eff}=\frac{G_F}{\sqrt 2}V_{ij}V_{kl}^{*}[C_1(\mu)O_1(\mu)+ 
C_2(\mu)O_2(\mu)]\;,
\label{eff}
\end{eqnarray}
with the four-fermion operators $O_1=({\bar q}_l q_k)({\bar q}_j q_i)$ and 
$O_2=({\bar q}_j q_k)({\bar q}_l q_i)$. The matching conditions 
of the Wilson coefficients are given by
$C_1(M_W)=1$ and $C_2(M_W)=0$.

The most widely adopted approach to exclusive nonleptonic heavy meson
decays is the BSW model \cite{BSW}, in which the factorization hypothesis
on the matrix elements of the operators $O_{1,2}$ is assumed. In this model
decay rates are expressed in terms of various hadronic transition form
factors. Employing Fierz transformation, the coefficient of the form
factors corresponding to external $W$ boson emissions is $a_1=C_1+C_2/N$,
and that corresponding to internal $W$ boson emissions is $a_2=C_2+C_1/N$.
The form factors may be related to each
other by heavy quark symmetry, and parametrized by different ansatz.
Nonfactorizable contributions, which can not be expressed in terms of
hadronic transition form factors, and nonspectator contributions from
$W$ boson exchanges are neglected.

Physical quantities such as decay amplitudes should not
depend on the renormalization scale $\mu$.
In principle, the matrix elements of the
four-fermion operators contain a $\mu$ dependence, which exactly cancels 
that of the  Wilson coefficients. In the BSW method, however,
nonleptonic matrix elements are factorized into two
matrix elements of (axial) vector currents. Since the currents are conserved,
the matrix elements have no anomalous scale dependence. Therefore, 
predictions from the BSW model are $\mu$-dependent, and can not be
physical. In the PQCD approach, the RG evolutions from $M_W$ to $t$ and
from $t$ to $1/b$ are taken into account, such that the factorization
formulas are scale-independent.

To circumvent the scale dependence in the BSW model, the Wilson
coefficients $a_i$ are
regarded as free parameters, and are determined by experimental data 
\cite{BSW}. However, the evaluation of the  hadronic form factors usually
involve some ansatz \cite{CT}, so that  the extraction of $a_1$ and $a_2$
is model dependent. On the other hand, it was
found that the ratio $a_2/a_1$ from an individual fit to the CLEO data of 
$B\to D^{(*)}\pi(\rho)$ \cite{A} varies significantly \cite{CT}. It was 
also shown that an allowed domain $(a_1,a_2)$ exists for the three classes 
of decays ${\bar B}^0\to D^{(*)+}$, ${\bar B}^0\to D^{(*)0}$ and 
$B^-\to D^{(*)0}$, only when the experimental errors are expanded to a 
large extent \cite{GKKP}. 
In the PQCD approach the Wilson coefficients, with their evolutions being
determined by RG equations as shown in Eqs.~(\ref{mb})-(\ref{mf}), are not
free universal parameters. The hard scale $t$
depends on meson dynamics, and is thus process-dependent. The specific
dynamics involved in different $B$ meson decays are then reflected by the
scale $t$, or equivalently, by the RG evolutions.

The BSW model encounters other difficulties. It has been known
that the large $N$ limit of $a_{1,2}$, {\it i.e.}, the choice 
$a_{1}=C_{1}(M_c)\approx 1.26$ and $a_{2}=C_{2}(M_c)\approx -0.52$,
with $M_c$ the $c$ quark mass, explains the data of charm decays \cite{BSW}. 
However, the same large $N$ limit of $a_{1}=C_{1}(M_b)\approx 1.12$ and 
$a_{2}=C_{2}(M_b)\approx -0.26$, $M_b$ being the $b$ quark mass, does not 
apply to the bottom case. That is, the different mechanism between
charm and bottom decays can not be understood in the BSW approach.
To overcome this difficulty, parameters $\chi$,
denoting the corrections from the nonfactorizable contributions, 
have been introduced \cite{C}. They lead to the effective coefficients
\begin{equation}
a^{\rm eff}_1=C_1+C_2\left(\frac{1}{N}+\chi_1\right)\;,\;\;\;\;
a^{\rm eff}_2=C_2+C_1\left(\frac{1}{N}+\chi_2\right)\;.
\end{equation}
$\chi$ should be negative for charm decays, canceling the color-suppressed 
term $1/N$, and be positive for bottom decays in order to enhance the 
predictions. 
In the PQCD formalism the nonfactorizable contributions do not lead to
additional parameters. They can be evaluated by convoluting the
nonfactorizable hard subamplitude (calculable in perturbation theory)
with the same meson wave functions as those for factorizable contributions
because of the universality. 
Using the three-scale factorization formulas, the mechanism responsible
for the sign change of $\chi$ has been found \cite{LT}.

\acknowledgments
This work was supported by the National Science Council of R.O.C.
under the Grant No. NSC-88-2112-M-006-013.

\end{document}